\documentclass[a4paper,manyauthors,nocleardouble,COMPASS]{cernphprep}

\usepackage{amssymb}
\usepackage{amsmath}
\usepackage{amsbsy}
\usepackage{xspace}
\usepackage{graphicx}
\usepackage{color}
\usepackage{colordvi}
\usepackage{cite} 
\usepackage{multicol}
\usepackage{booktabs}
\usepackage{multirow}
\usepackage{lscape}
\usepackage{bm}
\usepackage{epsfig} 

\RequirePackage[T1]{fontenc}
\newcommand{\eg}{{\it e.g.}\xspace}
\newcommand{\ie}{{\it i.e.}\xspace}

\begin{document}


\begin{titlepage}

\PHnumber{2015--328}
\PHdate{
\vspace*{-1.1\baselineskip}
\begin{tabular}[c]{r@{}}
February 16, 2015\\
rev. Feb. 21, 2016
\end{tabular}
}

\title{ 
Leading-order determination of the gluon polarisation
from semi-inclusive deep inelastic scattering data } 

\Collaboration{The COMPASS Collaboration}
\ShortAuthor{The COMPASS Collaboration}

\begin{abstract}
\label{abstract}

Using a novel analysis technique, the gluon polarisation in the nucleon is re-evaluated 
using the longitudinal double-spin asymmetry measured in the cross section of 
semi-inclusive single-hadron muoproduction with photon virtuality $Q^2>1~({\rm GeV}/c)^2$.
The data were obtained by the COMPASS experiment at CERN using a 160~GeV/$c$
polarised muon beam impinging on a polarised $^6$LiD target.  
By analysing the full range in hadron transverse momentum $p_{\rm T}$, 
the different $p_{\rm T}$-dependences of the underlying processes are separated using 
a neural-network approach.
In the absence of pQCD calculations at next-to-leading order in the selected kinematic domain,
 the gluon polarisation $\Delta g/g$ is evaluated at leading order in pQCD at a hard scale of 
$\mu^2= \langle Q^2 \rangle = 3 ({\rm GeV}/c)^2$. 
It is determined in three intervals of the nucleon momentum fraction carried by gluons, $x_{\rm g}$, 
covering the range $0.04 \!<\! x_{ \rm g}\! <\! 0.28$~ and does not exhibit a significant
dependence on $x_{\rm g}$. 
The average over the three intervals,
$\langle \Delta g/g \rangle = 0.113 \pm 0.038_{\rm (stat.)}\pm 0.036_{\rm (syst.)}$
at  $\langle x_{\rm g} \rangle \approx 0.10$, suggests that the gluon polarisation is positive
in the measured $x_{\rm g}$ range.
\end{abstract}

\vfill
\Submitted{(submitted to Eur. Phys. J. C)}

\end{titlepage}

{\pagestyle{empty}  
%
%
\section*{The COMPASS Collaboration}
\label{app:collab}
\renewcommand\labelenumi{\textsuperscript{\theenumi}~}
\renewcommand\theenumi{\arabic{enumi}}
\begin{flushleft}
C.~Adolph\Irefn{erlangen},
M.~Aghasyan\Irefn{triest_i},
R.~Akhunzyanov\Irefn{dubna}, 
M.G.~Alexeev\Irefn{turin_u},
G.D.~Alexeev\Irefn{dubna}, 
A.~Amoroso\Irefnn{turin_u}{turin_i},
V.~Andrieux\Irefn{saclay},
N.V.~Anfimov\Irefn{dubna}, 
V.~Anosov\Irefn{dubna}, 
W.~Augustyniak\Irefn{warsaw}
% ...
\end{flushleft}
%
%
\begin{Authlist}
\item [{\makebox[2mm][l]{\textsuperscript{\#}}}] Corresponding authors
\item \Idef{aveiro}{University of Aveiro, Department of Physics, 3810-193 Aveiro, Portugal} 
\item \Idef{bochum}{Universit\"at Bochum, Institut f\"ur Experimentalphysik, 44780 Bochum, Germany\Arefs{l}\Arefs{s}}
\item \Idef{bonniskp}{Universit\"at Bonn, Helmholtz-Institut f\"ur  Strahlen- und Kernphysik, 53115 Bonn, Germany\Arefs{l}}
\item \Idef{bonnpi}{Universit\"at Bonn, Physikalisches Institut, 53115 Bonn, Germany\Arefs{l}}
\item \Idef{brno}{Institute of Scientific Instruments, AS CR, 61264 Brno, Czech Republic\Arefs{m}}
\item \Idef{calcutta}{Matrivani Institute of Experimental Research \& Education, Calcutta-700 030, India\Arefs{n}}
\item \Idef{dubna}{Joint Institute for Nuclear Research, 141980 Dubna, Moscow region, Russia\Arefs{o}}
\item \Idef{erlangen}{Universit\"at Erlangen--N\"urnberg, Physikalisches Institut, 91054 Erlangen, Germany\Arefs{l}}
\item \Idef{freiburg}{Universit\"at Freiburg, Physikalisches Institut, 79104 Freiburg, Germany\Arefs{l}\Arefs{s}}
\item \Idef{cern}{CERN, 1211 Geneva 23, Switzerland}
\item \Idef{liberec}{Technical University in Liberec, 46117 Liberec, Czech Republic\Arefs{m}}
\item \Idef{lisbon}{LIP, 1000-149 Lisbon, Portugal\Arefs{p}}
\item \Idef{mainz}{Universit\"at Mainz, Institut f\"ur Kernphysik, 55099 Mainz, Germany\Arefs{l}}
\item \Idef{miyazaki}{University of Miyazaki, Miyazaki 889-2192, Japan\Arefs{q}}
\item \Idef{moscowlpi}{Lebedev Physical Institute, 119991 Moscow, Russia}
\item \Idef{munichtu}{Technische Universit\"at M\"unchen, Physik Department, 85748 Garching, Germany\Arefs{l}\Arefs{r}}
\item \Idef{nagoya}{Nagoya University, 464 Nagoya, Japan\Arefs{q}}
\item \Idef{praguecu}{Charles University in Prague, Faculty of Mathematics and Physics, 18000 Prague, Czech Republic\Arefs{m}}
\item \Idef{praguectu}{Czech Technical University in Prague, 16636 Prague, Czech Republic\Arefs{m}}
\item \Idef{protvino}{State Scientific Center Institute for High Energy Physics of National Research Center `Kurchatov Institute', 142281 Protvino, Russia}
\item \Idef{saclay}{CEA IRFU/SPhN Saclay, 91191 Gif-sur-Yvette, France\Arefs{s}}
\item \Idef{taipei}{Academia Sinica, Institute of Physics, Taipei, 11529 Taiwan}
\item \Idef{telaviv}{Tel Aviv University, School of Physics and Astronomy, 69978 Tel Aviv, Israel\Arefs{t}}
\item \Idef{triest_u}{University of Trieste, Department of Physics, 34127 Trieste, Italy}
\item \Idef{triest_i}{Trieste Section of INFN, 34127 Trieste, Italy}
\item \Idef{triest_ictp}{Abdus Salam ICTP, 34151 Trieste, Italy}
\item \Idef{turin_u}{University of Turin, Department of Physics, 10125 Turin, Italy}
\item \Idef{turin_i}{Torino Section of INFN, 10125 Turin, Italy}
\item \Idef{illinois}{University of Illinois at Urbana-Champaign, Department of Physics, Urbana, IL 61801-3080, U.S.A.}   
\item \Idef{warsaw}{National Centre for Nuclear Research, 00-681 Warsaw, Poland\Arefs{u} }
\item \Idef{warsawu}{University of Warsaw, Faculty of Physics, 02-093 Warsaw, Poland\Arefs{u} }
\item \Idef{warsawtu}{Warsaw University of Technology, Institute of Radioelectronics, 00-665 Warsaw, Poland\Arefs{u} }
\item \Idef{yamagata}{Yamagata University, Yamagata, 992-8510 Japan\Arefs{q} }
\end{Authlist}
%
%
\renewcommand\theenumi{\alph{enumi}}
\begin{Authlist}
\item [{\makebox[2mm][l]{\textsuperscript{*}}}] Deceased
\item \Adef{a}{Also at Instituto Superior T\'ecnico, Universidade de Lisboa, Lisbon, Portugal}
\item \Adef{b}{Also at Department of Physics, Pusan National University, Busan 609-735, Republic of Korea and at Physics Department, Brookhaven National Laboratory, Upton, NY 11973, U.S.A. }
\item \Adef{r}{Supported by the DFG cluster of excellence `Origin and Structure of the Universe' (www.universe-cluster.de)}
\item \Adef{d}{Also at Chubu University, Kasugai, Aichi, 487-8501 Japan\Arefs{q}}
\item \Adef{e}{Also at KEK, 1-1 Oho, Tsukuba, Ibaraki, 305-0801 Japan}
\item \Adef{g}{Also at Moscow Institute of Physics and Technology, Moscow Region, 141700, Russia}
\item \Adef{v}{Supported by Presidential grant NSh - 999.2014.2}
\item \Adef{ep}{Also at University of Eastern Piedmont, 15100 Alessandria, Italy}
\item \Adef{h}{Present address: RWTH Aachen University, III. Physikalisches Institut, 52056 Aachen, Germany}
\item \Adef{i}{Present address: Uppsala University, Box 516, SE-75120 Uppsala, Sweden}
\item \Adef{c}{Supported by the DFG Research Training Group Programme 1102  ``Physics at Hadron Accelerators''}
%
%
\item \Adef{l}{Supported by the German Bundesministerium f\"ur Bildung und Forschung}
\item \Adef{s}{Supported by EU FP7 (HadronPhysics3, Grant Agreement number 283286)}
\item \Adef{m}{Supported by Czech Republic MEYS Grant LG13031}
\item \Adef{n}{Supported by SAIL (CSR), Govt.\ of India}
\item \Adef{o}{Supported by CERN-RFBR Grant 12-02-91500}
\item \Adef{p}{\raggedright Supported by the Portuguese FCT - Funda\c{c}\~{a}o para a Ci\^{e}ncia e Tecnologia, COMPETE and QREN,
 Grants CERN/FP 109323/2009, 116376/2010, 123600/2011 and CERN/FIS-NUC/0017/2015}
\item \Adef{q}{Supported by the MEXT and the JSPS under the Grants No.18002006, No.20540299 and No.18540281; Daiko Foundation and Yamada Foundation}
\item \Adef{t}{Supported by the Israel Academy of Sciences and Humanities}
\item \Adef{u}{Supported by the Polish NCN Grant DEC-2011/01/M/ST2/02350}
\end{Authlist}

\clearpage
}

\maketitle

\section{Introduction} \label{sec:int}

The experimental observation by EMC~\cite{EMC}
that quark spins contribute only a small fraction to the 
spin of the nucleon initiated a lot of 
new developments in spin physics
(for a review see \eg Ref.~\citen{review}).
In order to investigate the origin of the nucleon spin, it is essential to also determine 
the contribution of gluons, $\Delta g$.
Information about this quantity can be obtained indirectly by studying scaling
violations in the spin-dependent structure function $g_1$
(see Refs.~\citen{qcd_smc,qcd_e155,qcd_compass,h2} and references therein) or directly by 
measurements of the gluon polarisation $\Delta g/g$ in polarised lepton-nucleon or
proton-proton interactions 
(see Refs.~\citen{smc_dgg, comp_hipt_lowq2,hermes_dgg_new,comp_hipt,comp_charm,star3,phenix5,phenix6,star5,phenix7,star7}).  
Indirect determinations of $\Delta g$ suffer from poor accuracy due to the limited kinematic range, 
in which the structure function $g_1$ is measured. 
Most recent direct determinations by fits performed in the context 
of perturbative Quantum Chromodynamics (pQCD)
at next-to-leading order (NLO) in the strong coupling constant~\cite{dssv, nnpdf}, 
which include proton-proton data from RHIC, suggest that
the gluon polarisation is positive in the measured range of 
the nucleon momentum fraction carried by gluons, $0.05<x_{\rm g}<0.20$~.  

\begin{figure}[hb]
\begin{center}
\includegraphics[width=0.7\textwidth]{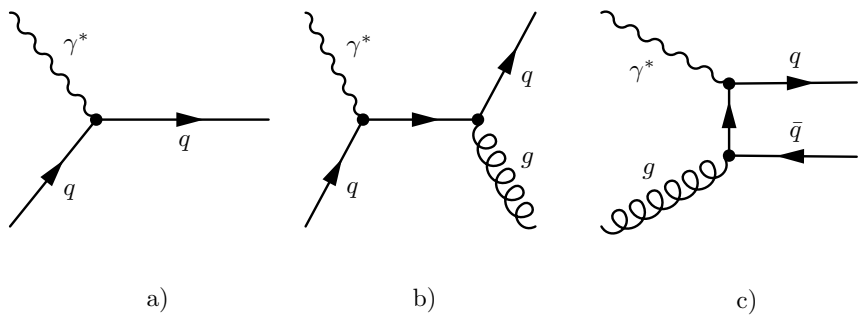}
\caption{Feynman diagrams for
a) the leading-order process (LP),
b) gluon radiation (QCDC), and 
c) photon--gluon fusion (PGF).}
\label{fig:pgf}
\end{center}
\end{figure}

In deep inelastic scattering (DIS),
the leading-order virtual-photon absorption process (LP) does not
provide direct access to the gluon distribution since the virtual photon does
not couple to the gluon. 
Therefore, higher-order processes
have to be studied, \ie QCD Compton scattering (QCDC) and Photon-Gluon Fusion (PGF), 
where only the latter is sensitive to the gluon helicity distribution.  
The diagrams for these two processes are shown in Fig.~\ref{fig:pgf}
together with that of the leading-order photon absorption process. 

In the leading-order process, the 
(small) transverse momentum of the produced hadron
originates from the intrinsic transverse momentum 
of the quark that was struck in the nucleon~\cite{kt} 
and the transverse momentum generated by the fragmentation of this quark. 
Here, transverse is meant relative to the virtual-photon direction 
in a frame where the nucleon momentum is parallel to this direction. 
The hard QCDC and PGF processes, on the contrary, can provide hadrons with high transverse momentum.  
Therefore, including in the analysis events with hadrons of large transverse 
momentum $p_{\rm T}$ enhances the contribution of higher-order processes. 
In earlier analyses, the contributions from LP and QCDC had to be subtracted in order 
to determine $\Delta g/g$ ~\cite{pt1}. A different approach is used 
in the present analysis, \ie a simultaneous extraction of 
$\Delta g/g$ and of the LP and QCDC asymmetries is performed  
using data that cover the full range in $p_{\rm T}$.
This ``all-$p_{\rm T}$ method'' takes advantage of the different $p_{\rm T}$-dependences
of the three processes in order to disentangle their contribution to the measured asymmetry.
Furthermore, this approach reduces systematic uncertainties with
respect to the one used previously~\cite{comp_hipt}.
In this paper, we re-analyse the semi-inclusive deep inelastic
scattering (SIDIS) data from COMPASS~\cite{comp_hipt},
applying the new all-$p_{\rm T}$ method.

\section{Experimental set-up and data sample} \label{sec:exp}

The COMPASS experiment is a fixed-target setup at the M2 beam line 
of the CERN SPS. The data used in this analysis were collected during four years: 2002 to 2004 and 2006.
For these measurements, longitudinally polarised positive muons were scattered
off a large polarised solid-state $^6$LiD target. 
A detailed description of the
experiment can be found elsewhere~\cite{nim}.  A major upgrade of the COMPASS
spectrometer was performed in 2005. For this analysis, the most relevant
improvement was a new target magnet that extended the angular acceptance
from $\pm 70$~mrad to  $\pm180$~mrad.

The average muon momentum was 160~GeV/$c$ and the average beam polarisation
was $\langle P_{\rm b} \rangle =-0.80\pm 0.04$.
The target material consisted of $^6$LiD beads
in a bath of $^3$He-$^4$He and was contained in
two target cells in 2002--2004 and in three cells in 2006.
The achieved target polarisation $P_{\rm t}$ 
was about $\pm$0.50 with a relative uncertainty of 5\%. 
Neighbouring target cells were polarised in opposite directions.
In order to cancel acceptance effects and to reduce systematic uncertainties,
the direction of the polarisation was reversed three times per day
in 2002--2004 and once per day in 2006.
The fact that not all nucleons in the target material are polarisable
is taken into account in the so-called effective dilution factor $f$.
It is given by the ratio of the total cross section
for muons on polarisable deuterons to the one on all nuclei 
taking into account their relative abundance in the target material.
Its value includes a correction factor 
$\rho = \sigma_{\rm d}^{1 \gamma} / \sigma_{\rm d}^{\rm tot}$
\cite{terad} accounting for radiative events on unpolarised deuterons and
a correction factor for the relative polarisation of deuterons bound
in $^6$Li compared to free deuterons. 
The dilution factor depends on the Bjorken scaling variable 
$x_{\rm Bj}$ and on the energy fraction $y$ carried by the exchanged virtual-photon; 
its average value for this analysis
is about 0.37 with a relative uncertainty of 5\%.

The data used for this analysis are selected by requiring an event to have
an interaction vertex located within the target fiducial volume.
An incoming and a scattered muon must be associated to this vertex. 
Moreover, the extrapolated trajectory of the incoming muon 
has to fully traverse
all target cells to assure that they all are exposed
to the same beam flux.
In order to select DIS events, the photon virtuality is required to be 
$Q^2>1~({\rm GeV}/c)^2$.
Events with $y < 0.1$ or $y > 0.9$ are rejected because the former are
more sensitive to time instabilities of the spectrometer, while the latter are
strongly affected by radiative effects.
With these $y$ limits, the squared invariant mass of the hadronic system,
$W^2$, is larger than $5~({\rm GeV}/c)^2$.
For a semi-inclusive single-hadron measurement, at least one charged
hadron has to be associated to the
vertex together with incoming and
scattered muons.
For the hadron with the highest $p_{\rm T}$, 
the requirement 0.05~GeV/$c$ $<p_{\rm T}<2.5$~GeV/$c$ 
has to be fulfilled. Here, the lower limit excludes electrons from $\gamma$   
conversion and the upper limit is discussed in Section~\ref{sec:mc}.
In order to suppress diffractive processes 
(mainly $\rho^{0}$ production), events are not accepted if they have
exactly two oppositely charged hadrons 
with $z_1 + z_2 > 0.95$, where $z_i$ is the energy fraction of hadron $i$ 
with respect to the energy of the virtual photon.

Compared to the previous analysis~\cite{comp_hipt}, there are two major
differences in the 
data selection process.  
First, at least one hadron instead of two hadrons is required in the final state.
Second, the smallest $p_{\rm T}$-value allowed 
for the hadron leading in $p_{\rm T}$ is 
lowered from 0.7~GeV/$c$ to 0.05~GeV/$c$.
After having applied all above described selection criteria, 
about 116 million events remain for the present analysis.

\section{Determination of $\Delta g/g$} \label{sec:form}

The predicted number of events $N^{\rm pre}(x_{\rm Bj})$ 
can be calculated from the
SIDIS cross sections of the three processes LP, QCDC, and PGF using 
the experimental acceptance $a$,
the number $n$ of scattering
centres in the target, the integrated beam flux $\Phi$,
and the unpolarised cross section $\sigma_0$ as

\begin{equation}
\begin{split}
N^{\rm pre}(x_{\rm Bj}) = a n \Phi  \sigma_0
\Big(1+ \big\langle fP_{\rm b} P_{\rm t} a_{\rm LL}^{\rm PGF} R_{\rm PGF} \; \, \frac{\Delta 
g}{g}(x_{\rm g})
\big\rangle + \big\langle fP_{\rm b}P_{\rm t} a_{\rm LL}^{\rm LP} R_{\rm LP} \;\,  A_1^{\rm 
LP}(x_{\rm Bj})
\big\rangle +\\
\big\langle fP_{\rm b}P_{\rm t} a_{\rm LL}^{\rm QCDC} R_{\rm QCDC} \;\, A_1^{\rm QCDC}(x_{\rm C}) 
\big\rangle
\Big)~.
\label{Eq:form:events}
\end{split}
\end{equation}

Here, the PGF part contains the gluon polarisation $\Delta g/g$. 
The two symbols $A_1^{\rm LP}$ and $A_1^{\rm QCDC}$ denote the same 
asymmetry;\footnote{They are also equal 
to  $A_1^{\rm LO}(x)$ in Eq.~(1) of Ref.~\citen{comp_hipt}}
the distinction is only kept to emphasise
the fact that in the new method there are two estimators of the same quantity.
This fact will be used in some systematic studies presented in Section \ref{sec:sys}.  
In Eq.~(\ref{Eq:form:events}), 
the predicted number of events depends only 
on the Bjorken scaling variable $x_{\rm Bj}$, as all other
variables are integrated over the experimental kinematic domain.
The label $i\in\{\rm LP, QCDC, PGF\}$ 
will be used to denote the three processes depicted in
Fig.~\ref{fig:pgf}. 
Each process has a characteristic nucleon momentum fraction:
$x_{\rm LP} \equiv x_{\rm Bj}$, $x_{\rm QCDC} \equiv x_{\rm C}$, 
$x_{\rm PGF} \equiv x_{\rm g}$.
For a given $x_{\rm Bj}$, the corresponding nucleon momentum fractions
carried by quarks in the QCDC process, $x_{\rm C}$, and by
gluons in the PGF process, $x_{\rm g}$, 
are in general larger, and their values depend on the kinematics of the event.
For each process $i$, the relative contribution is denoted by $R_i$
and the analysing power $a_{\rm LL}^i$ is given by
the asymmetry of the partonic cross section~\cite{poldis}.
The analysing power is proportional to the depolarisation factor $D$
that represents the fraction of the muon polarisation transferred to the virtual
photon, where for LP holds $a_{\rm LL}^{\rm LP}=D$.

Equation~(\ref{Eq:form:events})
is valid at leading order (LO) in pQCD
assuming spin-independent fragmentation.
A possible spin dependence of the fragmentation process~\cite{aram} 
can be neglected in the COMPASS kinematic region.
Equation~(\ref{Eq:form:events}) can be written in a more concise form as

\begin{equation}
 N^{\rm pre}(x_{\rm Bj})=\alpha \left(1+ \sum_i \left\langle \beta_i \;  A^i(x_i)
\right\rangle \right)~.
\label{Eq:form:events2}
\end{equation}

Here, $\alpha = a n \Phi  \sigma_0$, $\beta_i =fP_{\rm b} P_{\rm t}a_{\rm LL}^i R_i$ and
$\langle \beta_i  A^i(x_i) \rangle$ denotes the average
of $\beta_i  A^i(x_i)$ over the experimental kinematic domain.
For simplicity of notation, a possible $x_i$ dependence of $\beta_i$ is omitted in 
Eq.~(\ref{Eq:form:events2}).

The data were taken simultaneously for the upstream ($u$) and downstream
($d$) target cells, in which the material was polarised longitudinally in 
opposite directions. For the 2006 data taking, 
the label $u$ refers to the two outer cells and $d$ to the central cell. 
The directions of the polarisation 
were periodically 
reversed; the configuration 
before and after a reversal
will be denoted by $(u,d)$ and $(u',d')$, respectively.
For a stable apparatus it is expected that
$\alpha_u/\alpha_{d}=\alpha_{u'}/\alpha_{d'}$.
The data sample is divided into 40 periods,
over which the apparatus is indeed found to be stable.
Independent analyses are performed in each of these periods
and the final result is obtained as 
weighted average of the 40 single ones.

The gluon polarisation $\Delta g/g$ is evaluated 
using the set of four equations 
obtained from Eq.~(\ref{Eq:form:events}) for the 
four possible configurations of target cells and polarisation 
directions ($k=u$, $d$, $u'$, $d'$).
The process fractions $R_i$, the momentum fractions $x_{\rm C}$, $x_{\rm g}$, 
and the analysing powers 
$a^{\rm QCDC}_{\rm LL}$,
$a^{\rm PGF}_{\rm LL}$ are determined using 
Monte Carlo (MC) simulations.
In the previous analysis~\cite{comp_hipt}, the asymmetry $A_1^{\rm LP}$ was evaluated from the
inclusive lepton--nucleon asymmetry $A_{\rm LL}^{\rm incl}$ .
In this analysis, $A_1^{\rm LP}$ is extracted simultaneously with $\Delta g/g$
from the same data. 

The method applied here was introduced in Ref.~\citen{method}
and already used for a determination of the gluon polarisation
using open-charm events~\cite{comp_charm}.
Its main advantage is that it allows for an elegant and less CPU intensive way to obtain
near optimal statistical uncertainty (in the sense of Cramer-Rao bound \cite{CramerRao}) in a multidimensional
analysis.

In order to extract simultaneously the signal $\Delta g/g$ 
and the background asymmetries $A_1^{\rm LP}$ and $A_1^{\rm QCDC}$, 
the event yields are considered separately 
for the three processes $i$. Moreover, since $\Delta g/g$, $A_1^{\rm LP}$, and
$A_1^{\rm QCDC}$ are known to be $x_i$ dependent, the analysis is performed
in bins of the corresponding 
variable $x_i$, which are indexed by $m$.

For each configuration $k=u$, $d$, $u'$, $d'$ we calculate weighted 
`predicted' and `observed' event yields, $\mathcal{N}_{i_m,k}^{\rm pre}$ and 
$\mathcal{N}_{i_m,k}^{\rm obs}$, respectively.
Using the weight $w=fP_{\rm b}a_{\rm LL}R$,
the observed weighted yield of events for 
process $i$ in the $m^{th}$
bin of $x_i$
is given by summing the corresponding weights $w_{i,n}$:

\begin{equation}
\mathcal{N}_{i_m,k}^{\rm obs}= \sum_{n=1}^{N_k} \epsilon_{m,i} w_{i,n} = \sum_{n=1}^{N_k} 
 \epsilon_{m,i} f_n P_{b,n} a_{{\rm LL},n}^{i} R_{i,n} \,~.
\end{equation}

The sum runs over $N_k$, the number of events 
observed for configuration $k$,
and $\epsilon_{m,i}$ is equal to 1 if for a
given event its momentum fraction
$x_i$ falls into the $m^{th}$ bin, and zero otherwise.
The target polarisation is not included in the weight because its value changes 
with time.
Since one knows only the probabilities $R_i$ that the event originated
from a particular partonic process, each event contributes 
to all three event yields $\mathcal{N}_{{\rm PGF}_m,k}^{\rm obs} \;$, 
$\mathcal{N}_{{\rm QCDC}_{m'},k}^{\rm obs} \;$, and $\mathcal{N}_{ {\rm LP}_{m''},k}^{\rm obs}$.
The correlation between these events yields is taken into account 
by the covariance matrix 
$cov_{i_m j_{m'},k}=\sum_{n=1}^{N_k} \epsilon_{m,i} \epsilon_{m'\!,j} w_{i,n} w_{j,n}$.

The predicted weighted yield
of events of each type, $\mathcal{N}_{i_m,k}^{\rm pre}$,
is approximated by

\begin{equation}
\mathcal{N}_{i_m,k}^{\rm pre} \approx \alpha_{k,w_{i_m}} 
\left(1+ \sum_j \sum_{m'} \langle \beta_{j_{m'}}   \rangle_{w_{i_m}} \langle 
A^j(x_j)\rangle_{m'} 
\right)~, 
\end{equation}
where $\alpha_{k,w_{i_m}}$ is the weighted value of $\alpha_k$ 
and 
\begin{equation}
\langle \beta_{j_{m'}} \rangle_{w_{i_m}} \approx  
  \frac{
   \sum_{n=1}^{N_k} \epsilon_{m,i}   \epsilon_{m'\!,j} \beta_{j,n} w_{i,n}}
   {\sum_{n=1}^{N_k} \epsilon_{m,i} w_{i,n}}.
\end{equation}

Here, the above confirmed assumption 
$\alpha_{u,w_{i_m}}/ \alpha_{d,w_{i_m}}$ = $\alpha_{{u'},w_{i_m}}/ \alpha_{{d'},w_{i_m}}$ 
is used as well as the additional assumption  
$\langle \beta_j A^j(x_j) \rangle  \simeq   \langle \beta_j \rangle  \langle A^{j}(x_j) \rangle$.
Knowing the number of observed and predicted events as well as 
the covariance matrix, the standard definition of $\chi^2$ is used,
$ \chi^2= ( \pmb{\mathcal{N}}^{\rm \bf obs} -  \pmb{\mathcal{N}}^{\rm \bf pre})^ T 
cov^{-1} ( \pmb{\mathcal{N}}^{\rm \bf obs} -  \pmb{\mathcal{N}}^{\rm \bf pre})$,
where  $\pmb{\mathcal{N}}^{\rm \bf obs}$ and $\pmb{\mathcal{N}}^{\rm \bf pre}$ are vectors
with the components  $\mathcal{N}_{i_m,k}^{\rm obs}$ and $\mathcal{N}_{i_m,k}^{\rm pre}$, respectively. 
The values of $\Delta g/g$, $A_1^{\rm LP}$  and $A_1^{\rm QCDC}$ are obtained 
by minimisation of $\chi^2$ using the programme {\sc MINUIT}~\cite{minuit}.
The HESSE method from the same package is used to calculate the uncertainties.
In the present analysis we use 12 bins in $x_{\rm Bj}$, 6 in $x_{\rm C}$ and 
1 or 3 bins in $x_{\rm g}$.
In the COMPASS kinematic region holds $x_{\rm C} \gtrapprox 0.06$, so 
that 
the same binning can be used for  $x_{\rm C}$ as for the six highest bins in $x_{\rm Bj}$.
In order to further constrain $\Delta g/g$, one can 
eliminate several parameters from the fit by using the relation
$A_1^{\rm LP}(x) = A_1^{\rm QCDC}(x)$.
The presented equality does not hold for individual events, but only for 
classes of events, \ie there are LP events with $x_{\rm Bj}=0.10$ 
and there are QCDC events with $x_{\rm C}=0.10$, 
for which $x_{\rm Bj}$ is usually much smaller than 0.10~. Note that for a given event only the probability is known, to which class
it belongs. Hence even if the above equality is used in the analysis, any 
event will be still characterised by
different values of $x_{\rm Bj}$ and $x_{\rm C}$ in addition to $x_{\rm g}$. 

The data used for this analysis
is almost entirely dominated by the LP process, 
as the required lower limit for $p_T$ is as small as 0.05~GeV/$c$. 
It thus provides to the applied $\chi^2$ minimisation 
procedure enough lever-arm for a separation
between the LP and PGF processes, 
which allows for a simultaneous extraction of their asymmetries.
As a result, a significant  reduction of both statistical and systematic 
uncertainties is achieved when comparing to Ref.~\citen{comp_hipt}.
The proposed method was fully tested using MC data, with
given $A_1^{\rm LP}$ and $\Delta g/g$ as input parameters.

The presented method to extract $\Delta g/g$ is model dependent.
In order to facilitate possible future NLO analyses of $\Delta g/g$, 
we also calculate the model-independent 
longitudinal double-spin asymmetries 
in the cross section of semi-inclusively measured single-hadron muoproduction,
$A^{\rm h}_{\rm LL}$.
They are extracted in bins of $x_{\rm Bj}$ and $p_{\rm T}$ 
of the hadron leading in $p_{\rm T}$ and are available in \ref{App:AppendixA}
We note that these asymmetries are not used directly 
in the all-$p_{\rm T}$ method presented in this paper.

\section{Monte Carlo Simulation and Neural Network Training}\label{sec:mc}

The DIS dedicated LEPTO event generator~\cite{lepto} (version 6.5)
is used to generate Monte Carlo (MC) events using the unpolarised 
cross sections of the three processes involved.
A possible contribution from resolved photon processes, not described in LEPTO,
is small in \cite{comp_hipt} and hence neglected.

The generated events are processed by the detector simulation
programme COMGEANT (based on GEANT3) 
and reconstructed in the same way as real events 
by the reconstruction programme CORAL.
The same data selection is then applied to real and MC events.
In Ref.~\citen{comp_ptlowq2} it was found that
simulations with the two hadron-shower models available in GEANT3, i.e. 
GHEISHA and FLUKA, give inconsistent results in the high-$p_{\rm T}$ region.
Hence events are included in the present analysis only, 
if the hadron leading in $p_{\rm T}$ has $p_{\rm T}$ $<2.5$~GeV/$c$.

\begin{figure}
\centerline{
\includegraphics[clip,width=1.0\textwidth]{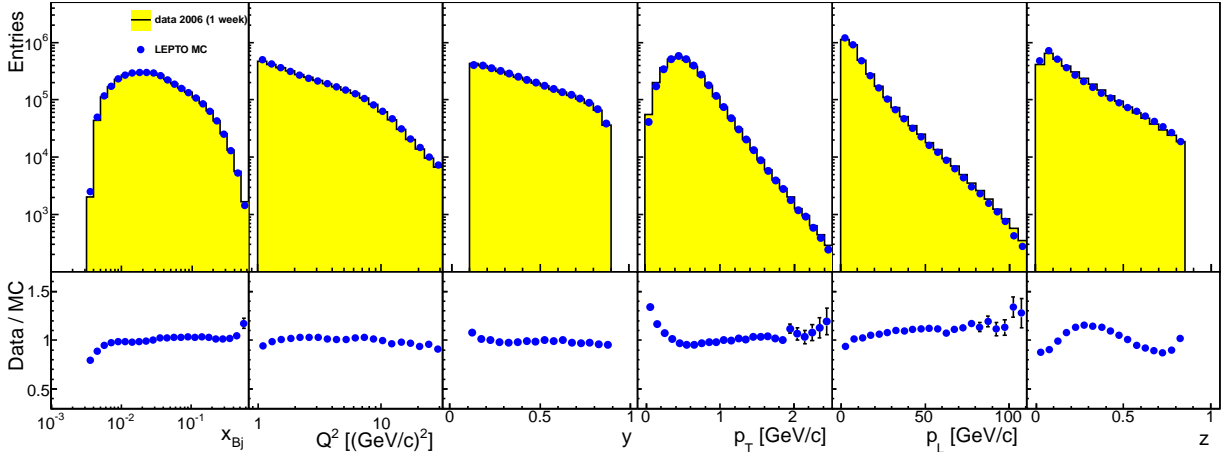}}
\caption{Comparison of kinematic distributions from data 
and MC simulations (top panels) and their
ratio (bottom panels) for the lepton variables
$x_{\rm Bj}$, $Q^2$, $y$
and for $p_{\rm T}$, $p_{\rm L}$ and $z$ of the hadron leading 
in $p_{\rm T}$, normalised to the number of events.}
\label{fig:mc:had}
\end{figure}

The best description of the data in terms of data-to-MC ratios for  
kinematic variables is obtained when using LEPTO with
the parton shower mechanism switched on, the fragmentation-function tuning 
as described in Ref.~\citen{comp_hipt}, and the PDF set 
of MSTW08LO$_{\rm 3fl}$ from Ref.~\citen{pdf_mstw08} 
together with the $F_L$-function option from LEPTO.
A correction for radiative
effects as described in Ref.~\citen{terad} is applied.
In Figure~\ref{fig:mc:had}, real
and MC data are compared for the lepton variables
$x_{\rm Bj}$, $Q^2$, $y$
and for $p_{\rm T}$, $p_{\rm L}$ and $z$ of the hadron leading in $p_{\rm T}$.
Here, $p_L$ denotes the longitudinal component of the hadron momentum.
The Monte Carlo simulation describes the data reasonably well over 
the full phase space. The largest discrepancy is observed
for low values of $p_{\rm T}$, where the LP process is dominant
so that this region has only limited impact on the extracted $\Delta g/g$ value.
The best description of the data in terms of data-to-MC ratios 
is the reason to select the 
above described MC sample for the extraction of the final $\Delta g/g$ value. 

\begin{figure}[tbp]
\centerline{\includegraphics[width=0.75 \textwidth]{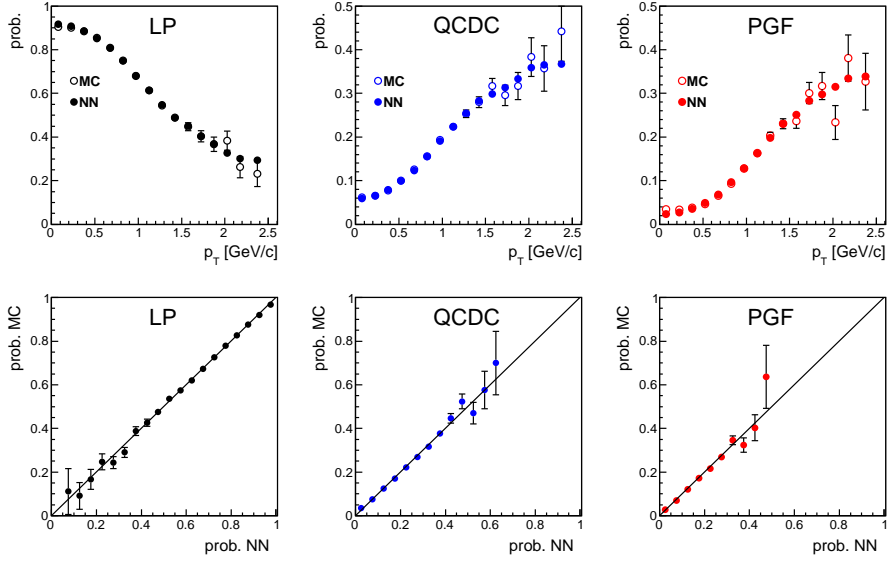}}
\caption{Top panels: Values of $R_{\rm LP}$, $R_{\rm QCDC}$, $R_{\rm PGF}$ 
obtained from MC and NN as a function of $p_{\rm T}$. Bottom panels:
MC probabilities in bins of NN probabilities.}
\label{fig:NN:stab}
\end{figure}

For a given set of input parameters, a neural network (NN) is trained to 
yield the corresponding
expectation values for 
the process fractions $R_i$, 
the momentum fractions $x_i$ and the analysing powers $a^i_{\rm LL}$.
The input parameter space is defined by $x_{\rm Bj}$, $Q^2$ and by $p_{\rm L}$, $p_{\rm T}$
of the hadron leading in $p_{\rm T}$.
The {\sc NETMAKER } tool kit  from Ref.~\citen{robert_NN} 
is used in
the analysis.\footnote{A feed-forward multi-layer perceptron
neural network is selected with the cost function 
defined by the mean squared difference between
expected output value and its neural network parametrisation.}
In the case that a clear distinction between the `true' MC value and
its NN parametrisation is needed,
for the latter one the superscript `{\rm NN}' will be added to the symbol denoting this 
variable, \eg $x_{\rm g}^{\rm NN}$. 
An example of the quality of the NN parametrisation is given in
the top panels of Fig.~\ref{fig:NN:stab}. It shows `true'
probabilities for LP, QCDC and PGF events as a function of $p_{\rm T}$
and the NN probabilities obtained for the same MC data.
While the LP probability falls with increasing $p_{\rm T}$, 
QCDC and PGF probabilities rise with comparable strength.
Another NN quality test is presented in the bottom panels of Fig.~\ref{fig:NN:stab},
where MC samples 
are selected in bins of the $R_i$ values returned by the NN,
which corresponds to the probability that the given event is of the 
process type $i$.
Using the true MC information, it is possible to verify
the generated fraction of each process
$i$ in the selected samples. 
A very good correlation is visible between NN output and the true MC composition.

\section{Systematic Studies }  \label{sec:sys}

With respect to the analysis method used in Ref.~\citen{comp_hipt},
two contributions to the systematic uncertainty are eliminated,
\ie the one related to the $x_{\rm C}$ approximation\footnote{\ie  $x_{\rm C}=x_{\rm C'}$ in Eq.~(3) 
of Ref.~\cite{comp_hipt}.}
and the one related to the parametrisation of $A_{1,{\rm d}}^{\rm incl}$.
The former approximation is simply not present in the current
method of $\Delta g/g$ extraction, and the latter input is not needed as 
$A^{\rm LP}$ is extracted from the same data set simultaneously with $\Delta g/g$.
The other major contributions to the 
total systematic
uncertainty are re-evaluated in the current analysis.
These are the limit on
experimental false asymmetries,
$\delta_{\rm false}$, the uncertainty related to the usage of MC in
the analysis, $\delta_{\rm MC}$, the impact of using a 
neural network to obtain the results, $\delta_{\rm NN}$,
and the uncertainty that is obtained by combining those of beam and target polarisations
and of the dilution factor, which is denoted as $\delta_{P_{\rm b}\!P_{\rm t}\!{\rm f}}$.
All these contributions to the systematic uncertainty are given in Table 
\ref{tab:sys} 
for the $\Delta g/g$ results obtained in
the full $x_{\rm g}$ range and for those obtained in three bins of $x^{\rm NN}_{\rm g}$.
The systematic uncertainty of the $\Delta g/g$ result, $\delta_{syst.}$, is calculated as
quadratic sum of the contributions $\delta_{\rm false}$, $\delta_{\rm MC}$,
$\delta_{\rm NN}$, and $\delta_{P_{\rm b}\!P_{\rm t}\!f}$.

\begin{table}[htb]
\begin{center}
\caption{Summary of contributions to the systematic uncertainty.} \label{tab:sys}
\begin{tabular}{ c c c c c c c}
\hline
 syst. unc.                   &\makebox[2em]{}& full $x_{\rm g}$ range &\makebox[3em]{}& 
$x^{\rm NN}_{\rm g}<0.10$ & $0.10<x^{\rm NN}_{\rm g}<0.15$ & $x^{\rm NN}_{\rm g}>0.15$ \\ \hline

$\delta_{\rm false}$&&   0.029  &&  0.039 & 0.022 & 0.014 \\ 
$\delta_{\rm MC}$  &&   0.017  &&  0.017 & 0.041 & 0.044 \\
$\delta_{\rm NN}$  &&   0.007  &&  0.007 & 0.007 & 0.018 \\
$\delta_{P_{\rm b}\!P_{\rm t}\!f}$   &&   0.010  &&  0.008 & 0.013 & 0.013 \\ \hline
$\delta_{\rm syst.}$&&  0.036  &&  0.044 & 0.049 & 0.051 \\ \hline
\end{tabular}
\end{center}
\end{table}

The false asymmetries are related to the stability of the spectrometer. 
The contribution of $\delta_{\rm false}=0.029$
is somewhat larger than that obtained in the previous analysis \cite{comp_hipt},
where it was additionally assumed that false
asymmetries are independent of $p_T$.\footnote{This assumption, 
when used in the current analysis,
would lead to a much lower value of $\delta_{\rm false}$ than previously.
This is due to the simultaneous extraction of $\Delta g/g$ and $A_1^{LP}$,
which are both affected by the same spectrometer instabilities,
thereby eliminating relative contributions to $\delta_{\rm false}$.}
The obtained uncertainty represents the difference between the 
final value of $\Delta g/g$
and the one obtained in a separate determination, in 
which the phase space region at low $x_{\rm Bj}$, low $p_{\rm T}$ and high $z$, which contributes to less than 5\% of the data sample, was removed from the analysis.
The values of $A^{h}_{\rm LL}$  obtained from this region are found to be different from those obtained in the main part of 
the phase space. From the detailed investigation of this discrepancy no clear conclusion could be drawn whether it is
a sign of an interesting physics effect appearing in this specific region of phase space, 
or it might be attributed to possible instabilities of the spectrometer. It appears worth noting that the
removal
of this specific phase space region from the analysis results in a value of $\Delta g/g$ that is 
larger by 0.029, albeit with very similar statistical uncertainties.

\begin{figure}[ht]
\begin{center}
\centerline{\includegraphics[width=0.49\textwidth]{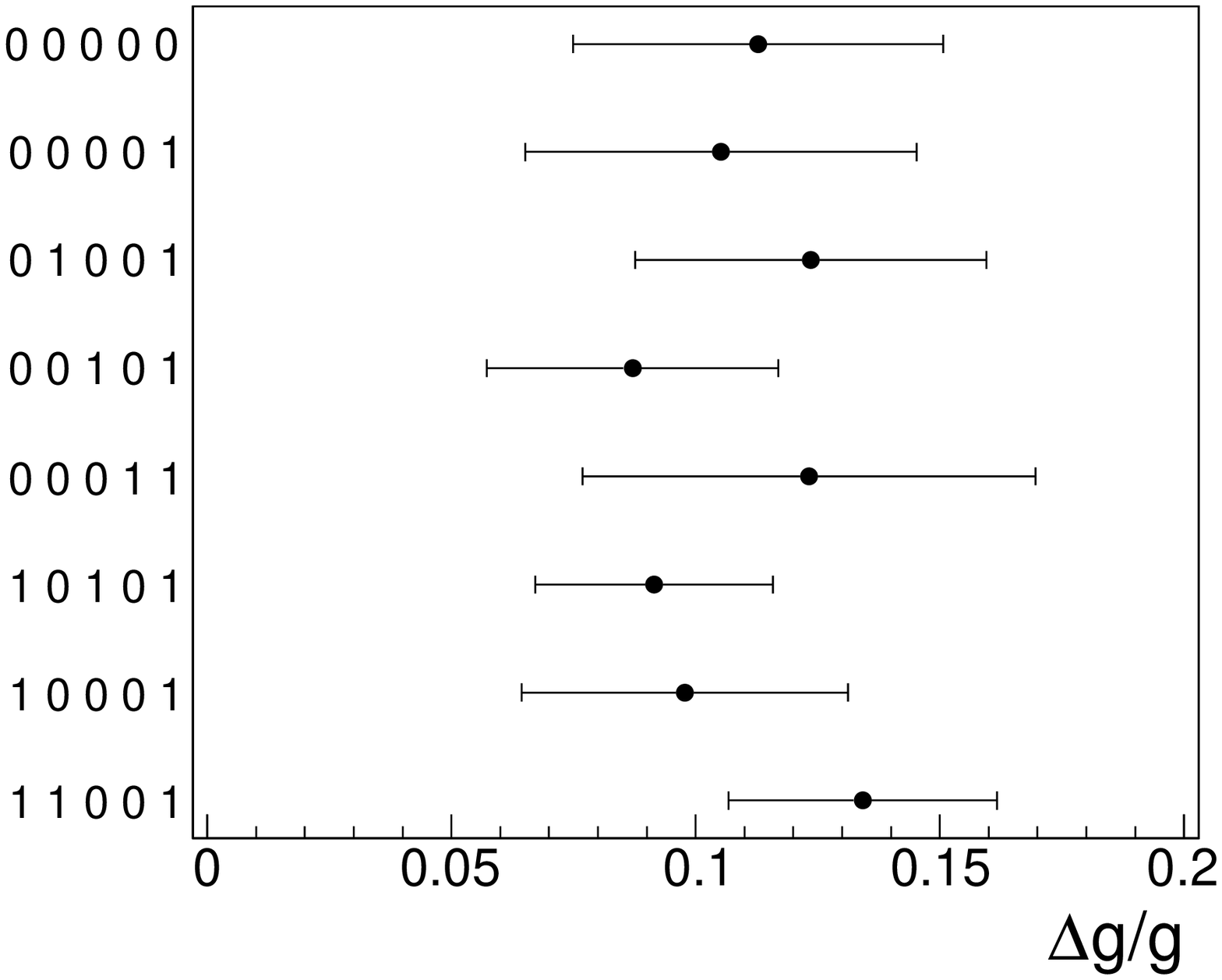}
            \includegraphics[width=0.49\textwidth]{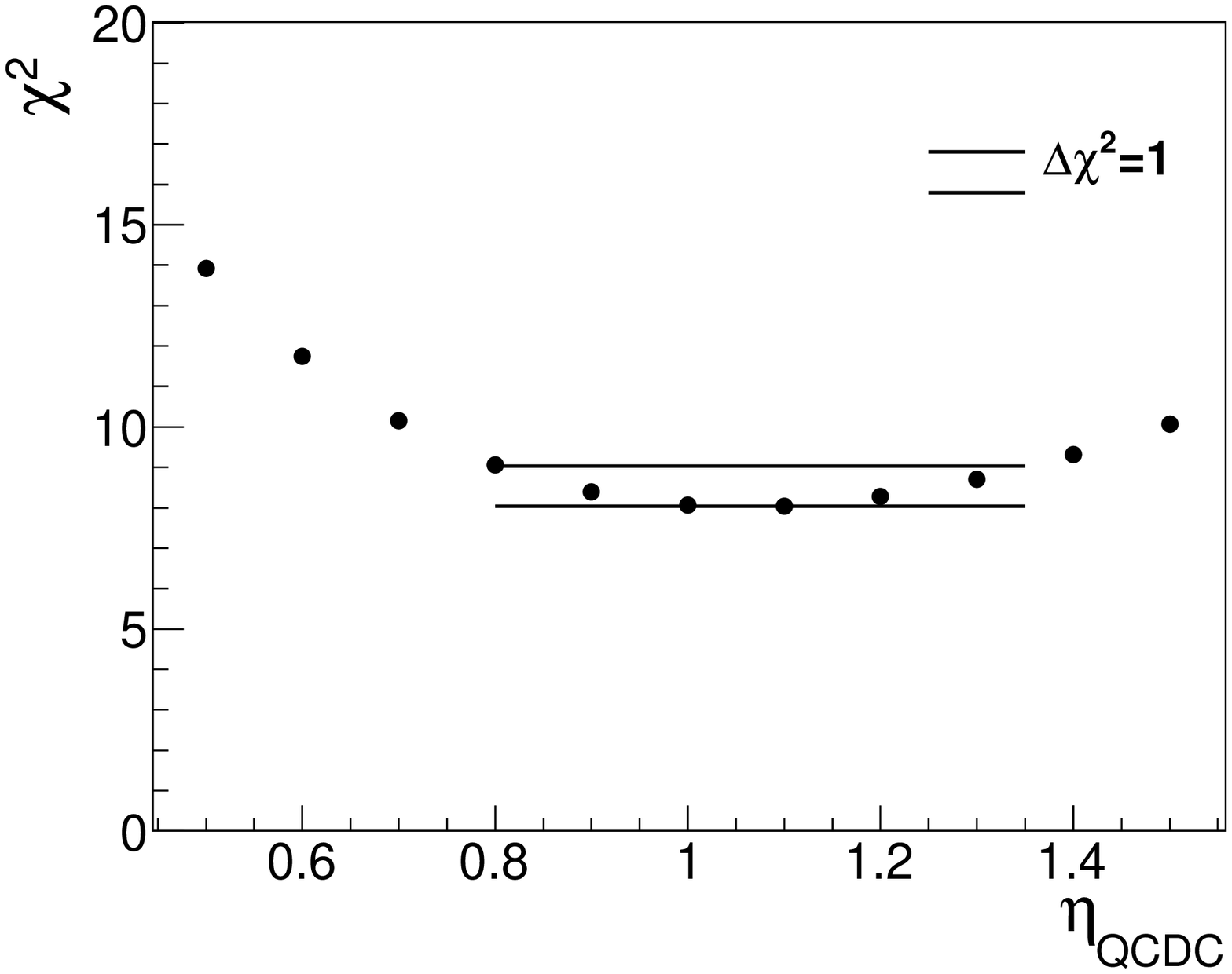}}
\caption{Left panel: Extracted values of $\Delta g/g$ and their statistical uncertainties
for eight different MC simulations.
A digit `1' at a certain position
in the 5-digit code shown on the ordinate means that the
corresponding simulation parameter was used differently as compared to
the code 00000 simulation that was used for the extraction of the final
$\Delta g/g$ results. The meaning of the digits is as follows (from left to
right): $1^{st}$: choice of the fragmentation functions
tuning; $2^{nd}$: usage of PS mechanism (here 0 means ON);  $3^{rd}$:
choice of PDF; $4^{th}$: usage of $F_L$ function from LEPTO (here 0 means ON);
$5^{th}$: choice of a program to simulate secondary interactions.
Right panel: The results of the $\chi^2$ scan of $\eta_{\rm QCDC}$, 
see text for details. }
\label{fig:sys1}
\end{center}
\end{figure}

Although the present analysis depends on the MC model used, the uncertainty
$\delta_{\rm MC}$ is found to be small.
It is evaluated by exploring the
parameter space of the model using eight different MC simulations.
These eight simulations 
differ by the tuning of the fragmentation functions
(COMPASS High-$p_{\rm T}$~\cite{comp_hipt} or LEPTO default), 
and by using or not using 
the
parton shower (PS) mechanism, which
also modifies the cut-off schemes used to
prevent divergences in the LEPTO cross-section calculation~\cite{lepto}.
Also, different PDF sets are used (MSTW08L or CTEQ5L~\cite{pdf_cteq5l}),
the longitudinal structure function $F_L$ from LEPTO is used or not used 
and alternatively FLUKA
or GEISHA is used for the simulation of secondary interactions.
Two observations are made when inspecting Fig.~\ref{fig:sys1}. 
The first one is that for the eight
different MC simulations the resulting values of $\Delta g/g$ are very similar; 
the root mean square (RMS) of the eight values, which is taken to
represent $\delta_{MC}$, amounts to only 0.017. 
The second observation is that the eight statistical uncertainties 
vary by up to a factor of two.

The explanation for the second observation is
that, in a good approximation,
the statistical uncertainty of $\Delta g/g$ is proportional to $1/R_{\rm 
PGF}$.
As in the eight different MC simulations the values
of $R_{\rm PGF}$ can vary by up to a factor two, 
large fluctuations of statistical uncertainties of $\Delta g/g$
are observed in Fig.~\ref{fig:sys1}.
The observation of a small RMS value can be understood by the following 
consideration. 
We start by using an equivalent of Eq.~(1) from Ref.~\citen{comp_hipt}, which is 
re-written for the one-hadron case.
Taking into account the experimental fact that 
the $A_{\rm LL}^{h}$ asymmetry weakly depends upon $p_{\rm T}$,
the left-hand side of the obtained equation
is effectively cancelled by the second term on the right-hand side,
which approximately corresponds to $A_{\rm LL}$ obtained in the low $p_{\rm T}$ region that is
dominated by LP. Under these assumptions $\Delta g/g$ is approximately given by

\begin{equation}  \label{Eq:dgg_simple}
\Delta g/g \approx - \frac{ a_{\rm LL}^{\rm QCDC} R_{\rm QCDC}}{a^{\rm PGF}_{\rm LL}
  R_{\rm PGF} } A_1^{\rm LP}(\langle x_{\rm C} \rangle \approx 0.14)~.
\end{equation}

The value of $A_1^{\rm LP}$ at $\langle x_{\rm C} \rangle=0.14$ is $\approx 0.087$, while
the value of $(a_{\rm LL}^{\rm QCDC} R_{\rm QCDC})/(a_{\rm LL}^{\rm PGF} R_{\rm PGF} )$ 
is $\approx 1.5$, 
resulting in $\Delta g/g \approx 0.13$. 
This value is not very different
from the result of the full analysis presented in Section \ref{sec:res}, 
which justifies the usage of Eq.~(\ref{Eq:dgg_simple}) for the  explanation of the small 
RMS. 
The values of $a_{\rm LL}^{\rm PGF}$ and $a_{\rm LL}^{\rm QCDC}$ in Eq.~(\ref{Eq:dgg_simple}) 
are quite stable with respect to the MC simulation used. As $a_{\rm LL}^{\rm PGF}$
depends mostly on $Q^2$ and $y$, which as inclusive variables 
are not affected by switching parton showers 
on or off nor by different fragmentation tunes, it is very similar 
in all eight MC simulations.
A similar consideration is valid for $a_{\rm LL}^{\rm QCDC}$, which depends mostly on $y$.
The ratio $R_{\rm QCDC}/R_{\rm PGF}$ is known more precisely than \eg the ratio 
$R_{\rm LP}/R_{\rm PGF}$ or $R_{\rm PGF}$ itself.%
\footnote{Note that the large instability of $R_{\rm PGF}$ itself
explains the large variation of the statistical uncertainty of $\Delta g/g$.}
One reason here is that both QCDC and PGF are treated in NLO, so that the strong 
coupling constant cancels in the 
cross-section ratio.
In addition, the hadron $p_{\rm T}$ in both
processes is dominated by the partonic cross section calculable in LO pQCD 
and not by the fragmentation process, for which the parameters were tuned.

The usage of 
a neural-network method leads to a 
systematic uncertainty $\delta_{\rm NN}=0.007$.
This uncertainty is estimated based on the
spread of $\Delta g/g$ values obtained from 
several NN parametrisations. These  parametrisations are obtained
by varying internal parameters of the NN training algorithm.

The relative systematic uncertainties of the beam and target
polarisation as well as of the dilution factor are estimated
to be 5\% each. Contrary to the method used in Ref.~\citen{comp_hipt},
in the all-$p_{\rm T}$ method the systematic uncertainty 
$\delta_{P_{\rm b}\!P_{\rm t}\!f}$
is proportional to the extracted value of $\Delta g/g$.
Therefore, it is evaluated to be 0.010~. 
The systematic uncertainties due to radiative corrections, 
due to the resolved-photon 
contribution, and due to remaining contributions from 
diffractive processes are estimated to be small and
can hence be safely neglected.

In the present analysis method,
$A_1^{\rm LP}$ and $A_1^{\rm QCDC}$ are two estimators
of the same quantity.
This fact allows us to perform additional consistency checks of the MC model
used in the analysis,
which were not possible in the analysis method used in Ref.~\citen{comp_hipt}.
The validity of the assumption $A_1^{\rm LP}(x)=A_1^{\rm QCDC}(x)$
can be verified by performing a standard $\chi^2$ test.
A possible failure of a $\chi^2$ test may 
indicate the use of incorrect $R_i$ and/or $a^i_{\rm LL}$ values 
in the analysis. 
This could happen if the MC tuning used in the analysis is wrong, 
or \eg higher-order corrections are substantial.
Such a consistency check was performed for all
eight MC samples, 
yielding a $\chi^2$ value between 3.9 and 13.1 for 6 degrees of
freedom. 
For the MC simulation used to obtain the quoted $\Delta g/g$ value, 
$\chi^2= 8.1$ was found, which means that 
the values of 
$A_1^{\rm QCDC}$ and $A_1^{\rm LP}$ are compatible.
Furthermore, one can also directly change the values of \eg
$a_{\rm LL}^{\rm QCDC} R_{\rm QCDC}$ obtained from NN,
and by checking the compatibility of the two $A_1$ values verify 
the consistency of data and MC model.
In the simplest test, we have added a
multiplicative factor $\eta_{\rm QCDC}$ to the MC value  of $a_{\rm LL}^{\rm QCDC} R_{\rm QCDC}$
and calculated the $\chi^2$ value of the compatibility test as a function
of $\eta_{\rm QCDC}$. As seen in the right panel of Fig.~\ref{fig:sys1},
the minimum value of $\chi^2$ is obtained for
$\eta_{\rm QCDC} \approx 1$, which supports the validity of the MC model.

The present analysis method assumes that $A_1^{\rm LP}$ and $\Delta g/g$ 
are independent of $p_{\rm T}$.
We have verified that if
different minimum  $p_{\rm T}$ cuts between 
0.05~GeV$/c$ and 1~GeV/$c$ 
are used in the data selection, 
the extracted values of $A_1^{\rm LP}$ and $\Delta g/g$ are compatible within statistical
uncertainties with the final results
when taking into account the correlations
between data samples. 
It is worth noting that 
this $p_{\rm T}$ scan in addition 
verifies 
that the removal of the region, in which the largest discrepancy 
between real and MC data is observed, 
does not bias the $\Delta g/g$ result. 
Similarly, in another test it was verified 
that compatible $\Delta g/g$ values are obtained  
with or without the cut $p_{\rm T}<2.5$~GeV/$c$.

\section{Results} \label{sec:res}

The re-evaluation of the gluon polarisation in the nucleon, yields
\begin{equation}
\langle \Delta g/g \rangle = 0.113 \pm 0.038_{(\rm stat.)} \pm 0.036_{(\rm syst.)},
\label{eq:dgg_val}\end{equation}
which is obtained at an average 
hard scale $\mu ^2 = \langle Q^2 \rangle  = 3$ (GeV/$c$)$^2$.
In the analysis, a correction is applied to account for the probability 
that the deuteron is in a D-wave state~\cite{d-wave}.
The presented value of the gluon polarisation was obtained 
assuming the equality of 
$A_1^{\rm LP}(x)$ and $A_1^{\rm QCDC}(x)$.
In the kinematic domain of the analysis, the average value of $x_{\rm g}$, 
weighted by $a_{\rm LL}^{\rm PGF} w_{\rm PGF}$,
is $\langle x_{\rm g} \rangle \approx 0.10$. 
In case $\Delta g/g$ can be approximated
by a linear function in the measured region of $x_{\rm g}$, the obtained
values of $\langle \Delta g/g \rangle$ correspond to 
the value of $\Delta g/g$ at this weighted average value of $x_{\rm g}$.
The obtained value of $\Delta g/g$
is positive in the measured $x_{\rm g}$ range and almost $3 \sigma_{\rm stat}$ from zero.
A similar conclusion is reached in
the NLO pQCD fits~\cite{dssv,nnpdf}, which include recent RHIC data.
The result of the present analysis 
agrees well with that of the previous one \cite{comp_hipt}, which 
was obtained from the same data ($\Delta g/g= 0.125 \pm 0.060 \pm 0.065$)~.
This comparison shows that the re-analysis using the new all-$p_{\rm T}$ 
method leads to a reduction of the 
statistical and systematic
uncertainty by a factor of 1.6 and 1.8, respectively.

The gluon polarisation is also determined in three 
bins of $x_{\rm g}^{\rm NN}$, which correspond to three ranges in $x_{\rm g}$. 
These ranges are partially overlapping due to 
an about 60\% correlation between $x_{\rm g}$ and $x^{\rm NN}_{\rm g}$,
which arises during the NN training.
The result on $\Delta g/g$ in three bins of $x_{\rm g}^{\rm NN}$ are
presented in Table \ref{tab:3res}. 
Within experimental uncertainties, 
the values do not show any significant dependence on $x_{\rm g}$.
Note that the events in the three bins of $x_{\rm g}^{\rm NN}$ are statistically independent.
In principle, for each $x_{\rm g}^{\rm NN}$ bin one could extract simultaneously 
$\Delta g/g$  and $A_1^{\rm LP}$ in 12 $x_{\rm Bj}$ bins, resulting in 36 
$A_1^{\rm LP}$ and three $\Delta g/g$ values. 
However, in order to minimise the statistical uncertainties of the obtained $\Delta g/g$ values, 
for a given $x_{\rm Bj}$ bin only one value of $A_1^{\rm LP}$ is extracted
instead of three.
As a result of such a procedure, a correlation between the three $\Delta g/g$ results 
may arise from the fit.
Indeed, a 30\% correlation is found between $\Delta g/g$ results obtained 
in the first and second $x_{\rm g}^{\rm NN}$ bins. 
The correlations of the results between the first 
or second and the third $x_{\rm g}^{\rm NN}$ bin
are found to be consistent with zero.

\begin{table}[htb]
\begin{center}
\caption{The values for $\langle \Delta g/g \rangle$ in  three $x_{\rm g}^{\rm NN}$ bins,
and for the full $x_g$ range. The $x_g$ range given in the third column 
corresponds to an interval in which 68\% of the MC events are found. }
\label{tab:3res}
\begin{tabular}{ r@{}l c c c }
\hline
$x^{\rm NN}_{\rm g} \;$& bin     & $\langle x_{\rm g} \rangle$ & $x_{\rm g}$ range & $\langle \Delta 
g/g \rangle$   \\ \hline
$0-$ & 0.10       & $0.08$ & $0.04-0.13$  & $0.087 \pm 0.050 \pm 0.044$  \\
$0.10-$& 0.15    & $0.12$ & $0.07-0.21$  & $0.149 \pm 0.051 \pm 0.049$  \\
$0.15-$& 1       & $0.19$ & $0.13-0.28$  & $0.154 \pm 0.122 \pm 0.051$   \\ \hline
$0-$& 1          & $0.10$ & $0.05-0.20$  & $0.113 \pm 0.038 \pm 0.036$   \\ \hline

\hline
\end{tabular}
\end{center}
\end{table}

\begin{figure}
\centerline{\includegraphics[width=0.5\textwidth]{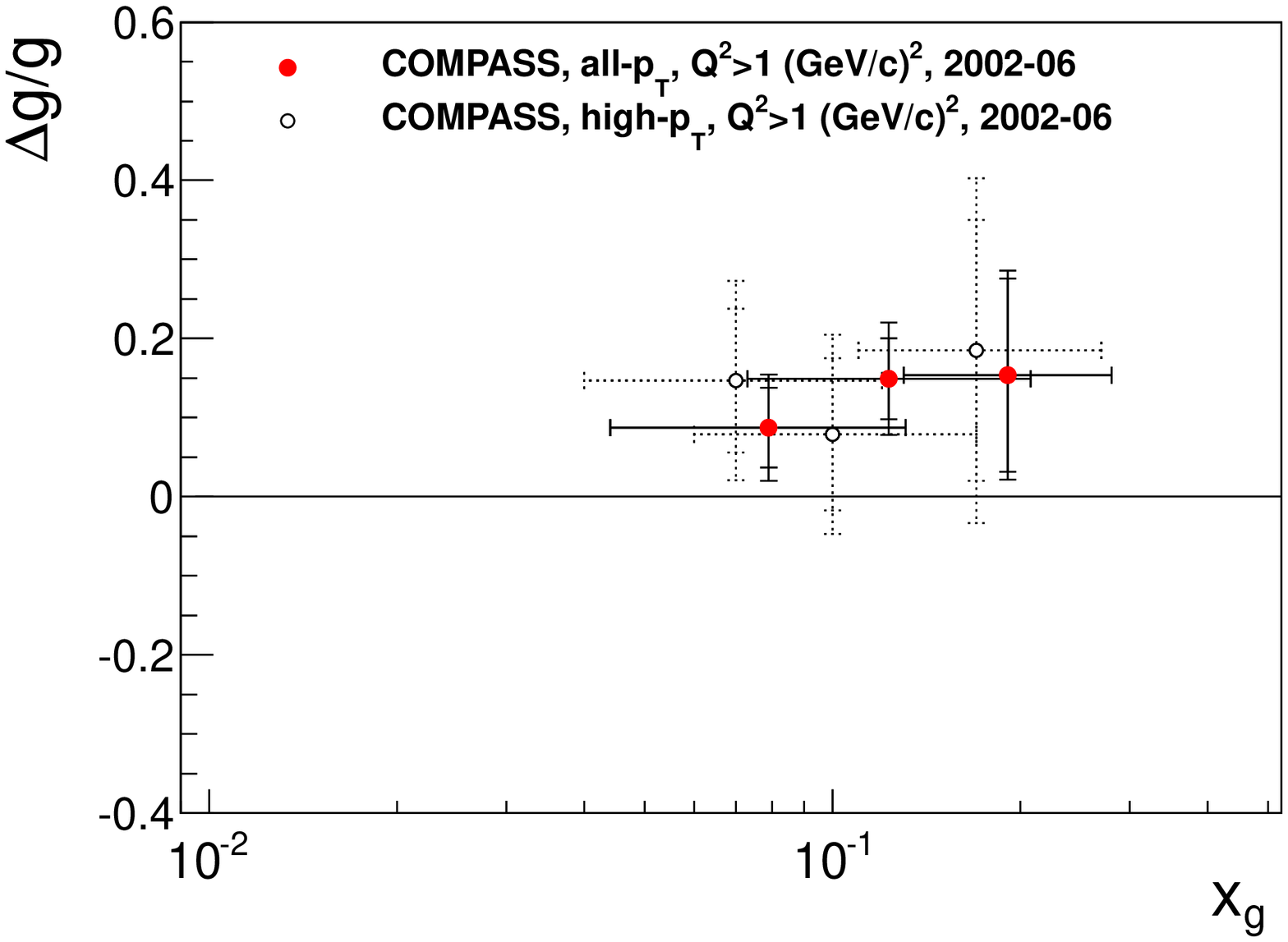}
\includegraphics[width=0.5\textwidth]{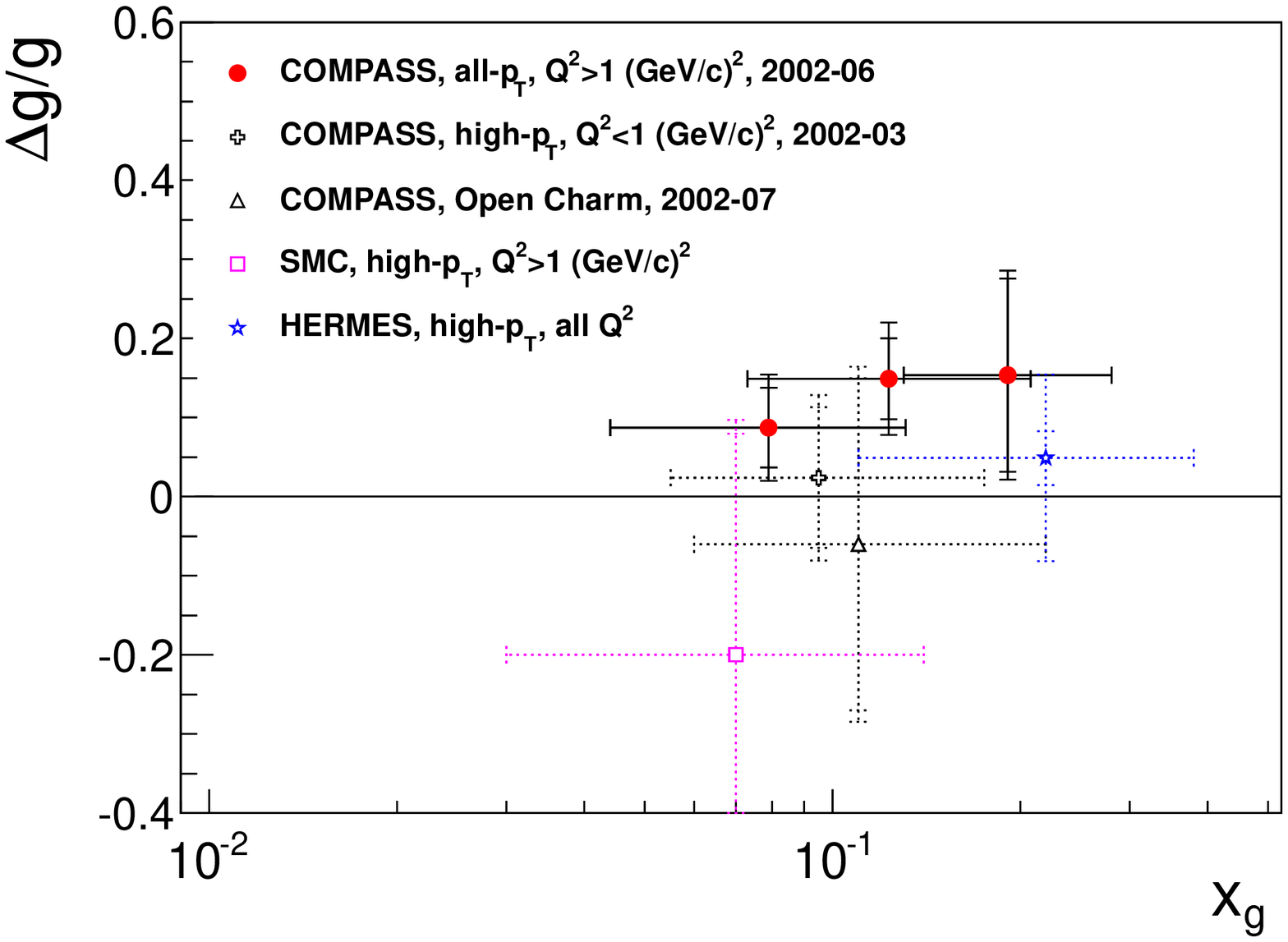} }
\caption{ The new results for $\Delta g/g$ in three $x_{\rm g}$ bins compared to
results of Ref.~\citen{comp_hipt} (left panel) and world data on
$\Delta g/g$ extracted in LO~\cite{smc_dgg,comp_hipt_lowq2, hermes_dgg_new, comp_charm} (right 
panel). The inner error bars represent the statistical uncertainties
and the outer ones the statistical and systematic uncertainties combined in quadrature. 
The horizontal
error bars represent the $x_{\rm g}$ interval in which 68\% of the MC events are found. }
\label{fig:dgg_old_new}
\end{figure}

A comparison of 
published \cite{comp_hipt} and present 
results is shown in the left panel of
Fig.~\ref{fig:dgg_old_new}. 
In addition to a clear reduction of the statistical uncertainties, 
a small shift in the average value of $x_{\rm g}$ is
observed, which originates from using slightly different data selection criteria in the
all-$p_{\rm T}$ analysis and also from differences between the two methods.
In the right
panel of Fig.~\ref{fig:dgg_old_new}, the new results are compared with the world
results on 
$\Delta g/g$ extracted in LO analyses~\cite{smc_dgg,comp_hipt_lowq2, hermes_dgg_new, 
comp_charm}, and good agreement is observed.
The new COMPASS results have the smallest combined
statistical and systematic uncertainty. 

\begin{figure}
\centerline{\includegraphics[width=0.5\textwidth]{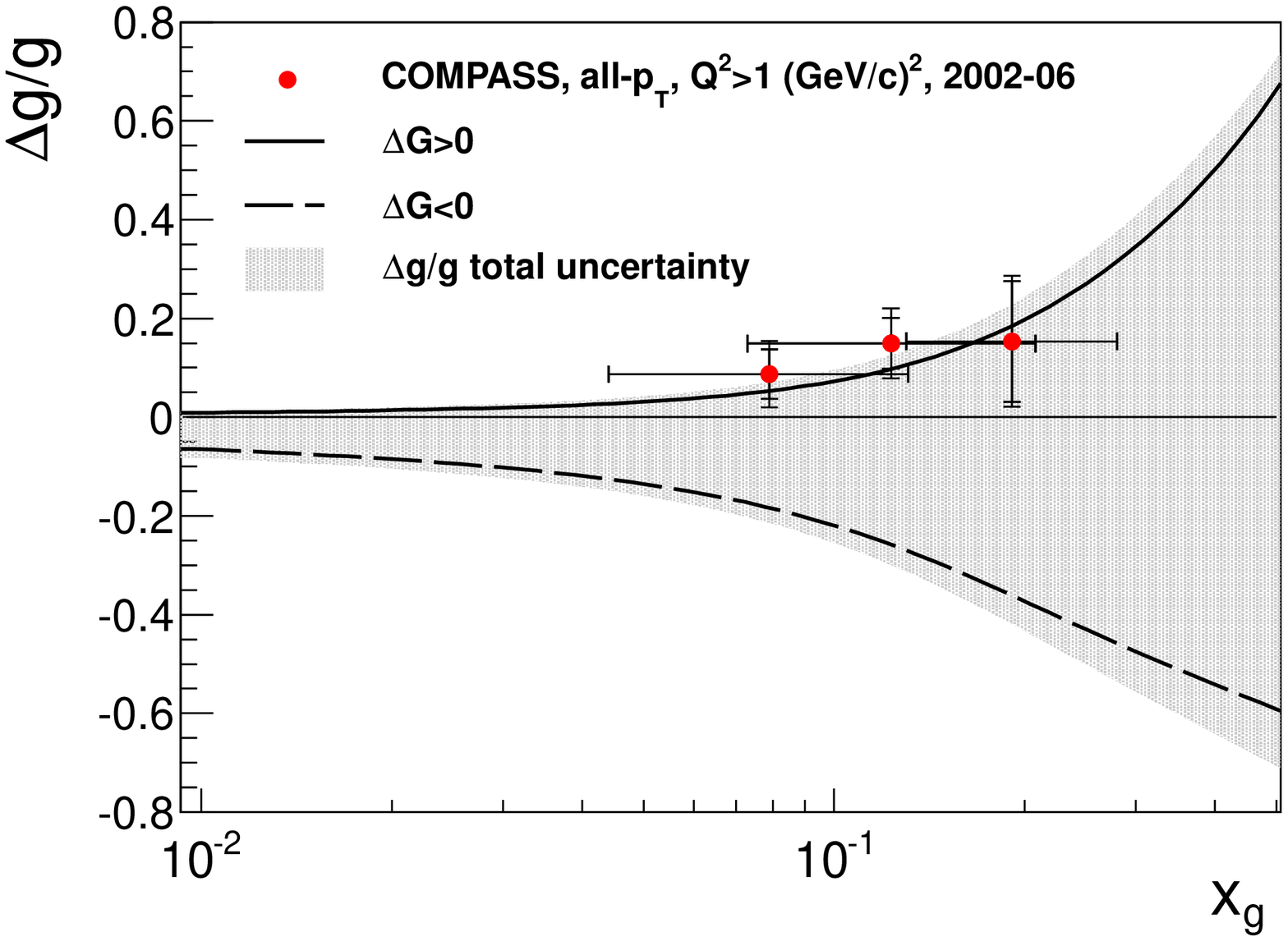}
\includegraphics[width=0.5\textwidth]{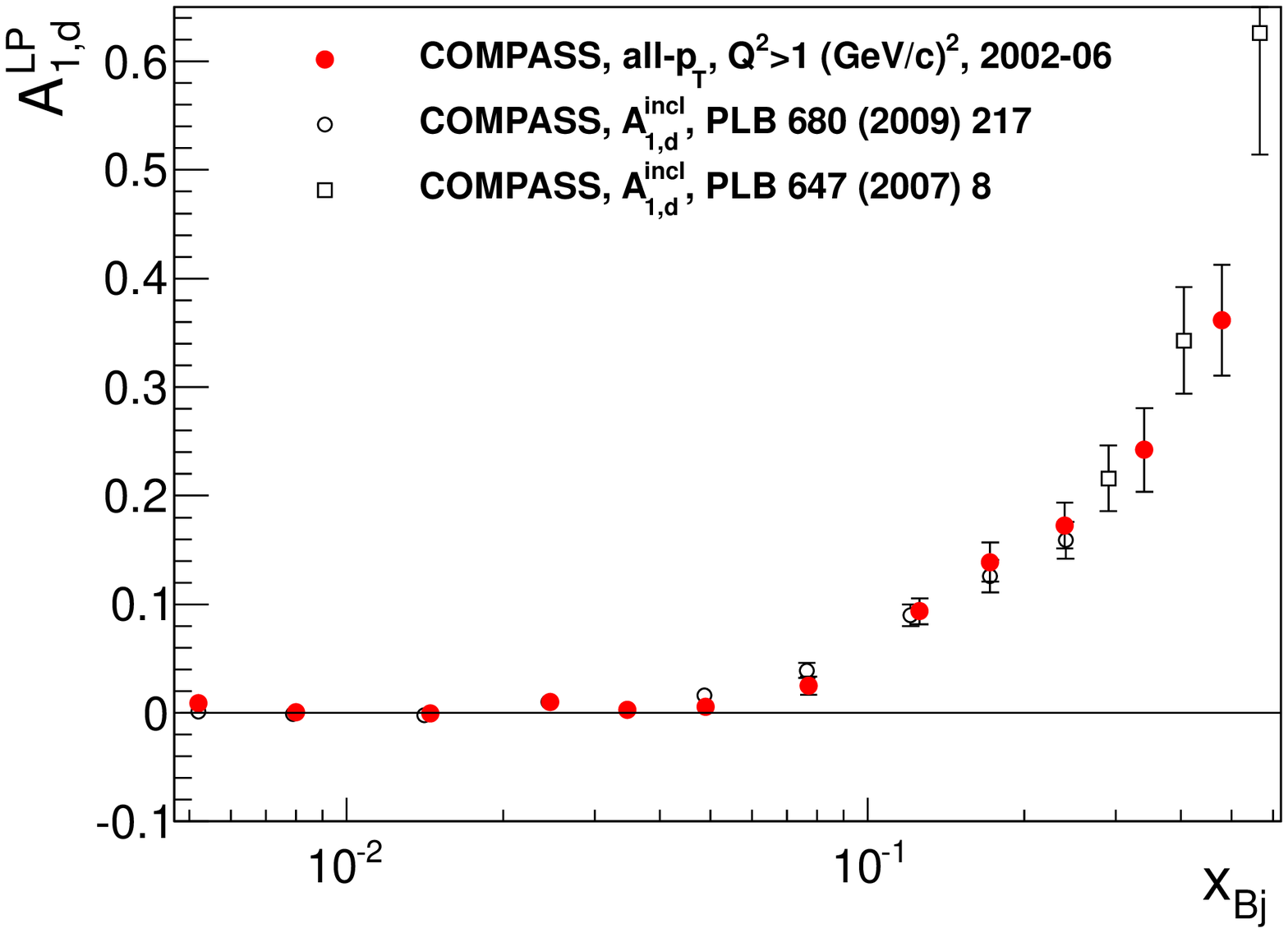}}
\caption{ Left panel:
Comparison of the 
LO results of the present analysis with the 
latest NLO QCD fit results from COMPASS~\cite{newg1_compass}.
Otherwise as in Fig.~\ref{fig:dgg_old_new}.
Right panel: Extracted values of $A_{1,{\rm d}}^{\rm LP}(x_{\rm Bj})$ and $A_{1,{\rm 
d}}^{\rm incl}$ 
from \cite{flav_compass,qcd_compass}. Here, only statistical uncertainties are shown.}
\label{fig:comp_fits}
\end{figure}

The left panel of Fig.~\ref{fig:comp_fits} shows the present results,
which are obtained at LO, in comparison to the most recent COMPASS
NLO $\Delta g/g$ parametrisation ~\cite{newg1_compass}.
The present results support solutions 
that yield positive values of $\Delta G$ 
in the NLO fit. 
Note that this comparison does not account for 
differences between LO and NLO
analyses.

For completeness, in the right panel of Fig.~\ref{fig:comp_fits} the extracted
values of $A_{1,{\rm d}}^{\rm LP}(x_{\rm Bj})$ are shown as full points.
They are consistent with zero at low $x_{\rm Bj}$ and
rise at higher $x_{\rm Bj}$.
The LP process measured in this analysis 
is the dominating contribution to the inclusive asymmetry 
$A_{1,{\rm d}}^{\rm incl}$, and the values of $A_{1,{\rm d}}^{\rm LP}$ and
$A_{1,{\rm d}}^{\rm incl}$ show very similar trends, as expected.  
The values of $A_{1,{\rm d}}^{\rm incl}$ for $x_{\rm Bj}<0.3$  are from 
Ref.~\citen{flav_compass}, while those for $x_{\rm Bj}>0.3$ are from
Ref.~\citen{qcd_compass}.

\section{Conclusions} \label{sec:sum}

Using COMPASS data on semi-inclusively measured single-hadron muoproduction off deuterium for a 
re-evaluation of the gluon
polarisation in the nucleon yields at LO in pQCD
$\langle \Delta g/g \rangle = 0.113 \pm 0.038_{\rm (stat.)} \pm 0.036_{\rm (syst.)}$ for a weighted average of
$\langle x_{\rm g} \rangle  \approx  0.10$ and an average
hard scale of 3 (GeV/$c$)$^2$.
This result is compatible with and supersedes
our previous result
\cite{comp_hipt} obtained from the same $Q^2 > 1$ (GeV/$c$)$^2$ data.
It favours a positive gluon polarisation in the measured $x_{\rm g}$ range.
The novel `all-$p_{\rm T}$ method' employed in the present analysis leads to a 
considerable reduction of
both 
statistical and systematic uncertainties, which 
is due to the cancellation of some uncertainties 
in the simultaneous determination of $\Delta g/g$ and $A_{1,{\rm d}}^{\rm LP}$. 

\section*{Acknowledgements}

We gratefully acknowledge the support of the CERN management and staff and 
the skill
and effort of the technicians of our collaborating institutes. 
This work was made possible thanks to the financial support of our funding 
agencies.

\setcounter{section}{0}
\renewcommand*\thesection{Appendix \Alph{section}}

\setcounter{table}{0}
\renewcommand*\thetable{\Alph{section}.\arabic{table}}

\section{}

Using the same data sample as ​used for the
$\Delta g/g$ analysis, which is described in this paper,​
also the longitudinal double​-spin asymmetry $A^{\rm h}_{\rm LL}$ 
is evaluated in a two-dimensional $12 \times 5$ binning in $x_{\rm Bj}​$ and
the transverse momentum of the hadron leading in $p_{\rm T}$.
The same 12 $x_{\rm Bj}$ bins are chosen as used for the determination of
$A_{1}^{\rm LP}$ in the main analysis.
As the contribution of higher-order processes increases with an 
increase of $p_{\rm T}$, this variable​ is chosen as the second one​.
The ​longitudinal double-spin asymmetries are​ extracted 
with the $2^{nd}$-order weighted method described in Ref.~\citen{compass_incl}
and shown in Table~\ref{tab:4app}.
In the selected 2-dimensional binning,
the systematic checks performed have shown
no presence of systematic effects within statistical uncertaint​ies​.
As a result, the systematic uncertaint​ies
of the asymmetries presented in Table~\ref{tab:4app}
are​ smaller than the respective statistical one​s.
Note that these asymmetries are not directly used for
the extraction of $\Delta g/g$ that is
presented in this paper.

\begin{landscape}

\begin{table}[htb]
\begin{center}
\caption{The values for $A^{\rm h}_{\rm LL}$ in bins of $x_{\rm Bj}$ and of $p_{\rm T}$ 
given in (GeV/$c$). } 
\label{tab:4app}
\begin{tabular}{ c c c r r r r r}
\hline

$x_{\rm Bj}$ range & $\langle x_{\rm Bj} \rangle$  &  $\langle Q^2 \rangle$ (GeV/$c)^2$ & 
\multicolumn{5}{c}{$A^{\rm h}_{\rm LL}$} \\ 
\hline

&  & &  $0.05<p_{\rm T}< 0.5$ & $0.5<p_{\rm T}< 1.0$ & $1.0<p_{\rm T}< 1.5$ &$1.5<p_{\rm T}< 2.0$ & $2.0<p_{\rm T}< 2.5$ \\ \hline

$0.003-0.006$&0.005  &1.2 & $0.0026  \pm 0.0046$& $0.0041 \pm 0.0051$ & $-0.005 \pm 0.013 $ &  $0.005  \pm 0.034 $ & $ -0.05   \pm	0.08 $\\
$0.006-0.010$&0.008 &1.4 & $-0.0020 \pm 0.0025$& $-0.0028\pm 0.0028$ & $-0.001 \pm 0.008 $ &  $0.004  \pm 0.020 $ & $ 0.01    \pm	0.05 $\\
$0.01-0.02$  &0.015 &1.8 & $-0.0013 \pm 0.0016$& $-0.0015\pm 0.0020$ & $-0.007 \pm 0.006 $ &  $0.000  \pm 0.016 $ & $ -0.03   \pm	0.04 $\\
$0.02-0.03$  &0.025 &2.3 & $0.0029  \pm 0.0019$& $0.0049 \pm 0.0026$ & $0.008  \pm 0.008 $ &  $0.016  \pm 0.024 $ & $ 0.07    \pm	0.06 $\\
$0.03-0.04$  &0.035 &2.8 & $0.0003  \pm 0.0023$& $0.0062 \pm 0.0034$ & $0.007  \pm 0.011 $ &  $0.051  \pm 0.033 $ & $ -0.03   \pm	0.09 $\\
$0.04-0.06$  &0.049 &3.8 & $0.0038  \pm 0.0022$& $0.0073 \pm 0.0033$ & $0.017  \pm 0.011 $ &  $-0.023 \pm 0.032 $ & $ 0.05    \pm	0.09 $\\
$0.06-0.10$  &0.077 &5.8 & $0.0062  \pm 0.0024$& $0.0117 \pm 0.0037$ & $0.013  \pm 0.012 $ &  $0.030  \pm 0.036 $ & $ 0.02    \pm	0.10 $\\
$0.10-0.15$  &0.12 &8.6 & $0.0204  \pm 0.0035$& $0.0214 \pm 0.0055$ & $0.037  \pm 0.018 $ &  $0.074  \pm 0.054 $ & $ 0.31    \pm	0.16 $\\
$0.15-0.20$  &0.17 &11.6 & $0.0282  \pm 0.0053$& $0.0368 \pm 0.0084$ & $0.027  \pm 0.027 $ &  $0.074  \pm 0.085 $ & $ -0.08   \pm	0.29 $\\
$0.20-0.30$  &0.24 &16.0 & $0.0439  \pm 0.0063$& $0.0414 \pm 0.0099$ & $0.114  \pm 0.032 $ &  $0.176  \pm 0.100 $ & $ -0.14   \pm	0.49 $\\
$0.30-0.40$  &0.34 &23.6 & $0.0696  \pm 0.0124$& $0.0690 \pm 0.0189$ & $-0.040 \pm 0.059 $ &  $0.056  \pm 0.199 $ & 	 \\ 
$0.40-1.00$  &0.48 &35.6 & $0.0822  \pm 0.0199$& $0.1154 \pm 0.0286$ & $0.076  \pm 0.078 $ &  $0.352  \pm 0.239 $ & 	 \\
\hline

\end{tabular}
\end{center}
\end{table}
\end{landscape}


\begin{thebibliography}{99}

\bibitem{EMC} EMC  Collaboration, J. Ashman {\it et al.},
Phys. Lett. B {\bf 206} (1988) 364; Nucl.\  Phys.\  B  {\bf 328} (1989) 1.

\bibitem{review} C.~A. Aidala, S.~D. Bass, D. Hasch and G.~K. Mallot,
Rev. Mod. \ Phys. {\bf 85} (2013) 655.


\bibitem{qcd_smc} SMC Collaboration, B. Adeva {\it et al.},
Phys. \ Rev.\ D  {\bf 58} (1998) 112002. 

\bibitem{qcd_e155}
E155 Collaboration, P.~L. Anthony {\it et al.},
Phys.\ Lett. B {\bf 493} (2000) 19.


\bibitem{qcd_compass} COMPASS Collaboration, V.~Yu. Alexakhin {\it et al.},
Phys.\ Lett.\ B  {\bf 647} (2007) 8.

\bibitem{h2} HERMES Collaboration,
A. Airapetian {\it et al.}, Phys.\ Rev. D  {\bf 75} (2007) 012007.



\bibitem{smc_dgg} SMC Collaboration, B. Adeva {\it et al.},
Phys. \ Rev. \ D {\bf 70} (2004) 012002.

\bibitem{comp_hipt_lowq2} COMPASS Collaboration,  E.~S. Ageev {\it et al.},
Phys.\ Lett.\ B {\bf 633} (2006) 25.

\bibitem{hermes_dgg_new}  HERMES Collaboration, A. Airapetian {\it et al.},
Journal of High Energy Physics {\bf 1008} (2010) 130.


\bibitem{comp_hipt}  COMPASS Collaboration, C. Adolph {\it et al.},
Phys.\ Lett.\ B {\bf 718} (2013) 922.

\bibitem{comp_charm} COMPASS Collaboration, C. Adolph {\it et al.},
Phys. \ Rev. \ D {\bf 87} (2013) 052018.



\bibitem{star3} STAR Collaboration, L. Adamczyk {\it et al.},  
Phys. \ Rev. \ D {\bf 89} (2014) 012001.

\bibitem{phenix5} PHENIX Collaboration, A. Adare {\it et al.},   
Phys. Rev. D {\bf 90} (2014) 012007.

\bibitem{phenix6}  PHENIX Collaboration, A. Adare {\it et al.},
Phys. \ Rev. \ D {\bf 84} (2011) 012006.


\bibitem{star5} STAR Collaboration, L. Adamczyk {\it et al.},
Phys. \ Rev. \ Lett. {\bf 115} (2015) 092002.

\bibitem{phenix7} PHENIX Collaboration, A. Adare {\it et al.},
Phys. \ Rev. \ Lett. {\bf 106} (2011) 062001.

\bibitem{star7} STAR Collaboration, L. Adamczyk {\it et al.},
Phys.\ Rev. \ Lett. {\bf 113} (2014) 072301. 


\bibitem{dssv} D. de Florian, R. Sassot, M. Stratmann and W. Vogelsang,
Phys.\ Rev. \ Lett. {\bf 113} (2014) 012001.

\bibitem{nnpdf}  E. R. Nocera, R. D. Ball, S. Forte, G. Ridolfi and J. Rojo, 
Nucl. \ Phys. B {\bf 887} (2014) 276.

\bibitem{kt} P. Renton and W.~S.~C. Williams, Ann. Rev. Nucl. Sci. {\bf 31} (1981) 193.

\bibitem{pt1} A. Bravar, D. von Harrach and A. Kotzinian,
Phys.\ Lett.\ B {\bf 421} (1998) 349.


\bibitem{nim} COMPASS Collaboration, P. Abbon  {\it et al.},
Nucl. Instrum. and Meth. A {\bf 577} (2007) 455.

\bibitem{terad}  A. A. Akhundov, D.Yu. Bardin, L. Kalinovskaya and T. Riemann, 
Fortschr. Phys. {\bf 44} (1996) 373.


\bibitem{poldis} A. Bravar, K. Kurek and R. Windmolders,
Comput.\ Phys.\ Commun.\ {\bf 105} (1997) 42.

\bibitem{aram} A. Kotzinian, Eur. \ Phys. \ J. \ C {\bf 44} (2005) 211.


\bibitem{method} J. Pretz and J.-M. Le Goff, 
Nucl.\ Instrum.\ Meth. A {\bf 602} (2009) 594.


\bibitem{CramerRao} H. Cramer, {\it Mathematical Methods of Statistics}, 
(Princeton Univ. Press, Princeton NJ, 1946); C. R. Rao,
Bulletin of the Calcutta Mathematical Society {\bf 37}, (1945) 81. 



\bibitem{minuit} F. James and M. Roos, Comput.\ Phys.\ Commun. {\bf 10} (1975) 343.



\bibitem{lepto} G. Ingelman, A. Edin and J. Rathsman,
Comput.\ Phys.\ Commun.\ {\bf 101} (1997)~108.


\bibitem{comp_ptlowq2} COMPASS Collaboration, C. Adolph {\it et al.},
Phys.\ Rev.\ D {\bf 88} (2013) 091101.



\bibitem{pdf_mstw08} A.~D. Martin, W.~J. Stirling, R.~S. Thorne and G. Watt,
Eur.\ Phys.\ J.\ C {\bf 64} (2009) 653.


\bibitem{robert_NN} R. Sulej, \emph{NetMaker}, http://www.ire.pw.edu.pl/$\sim$rsulej/NetMaker/.


\bibitem{pdf_cteq5l} CTEQ Collaboration, H.~L. Lai {\it et al.},
Eur.\ Phys.\ J.\ C {\bf 12} (2000) 375.


\bibitem{d-wave} R. Machleidt {\it et al.}, Phys. \ Rep. \ {\bf 149} (1987) 1.

\bibitem{newg1_compass} COMPASS Collaboration, C. Adolph {\it et al.},
subm. to Phys.\ Lett.\ B, hep-ex/1503.08935~.


\bibitem{flav_compass} COMPASS Collaboration, M. Alekseev {\it et al.}, 
Phys.\ Lett.\ B {\bf 680} (2009) 217.


\bibitem{compass_incl} COMPASS Collaboration, E.~S. Ageev {\it et al.},   
Phys.\ Lett.\ B {\bf 612} (2005) 154.



\end{thebibliography}
\end{document}